\documentstyle[jmb,epsfig,12pt,doublespace]{article} 
\begin{document}

\title{\textbf{The Folding Thermodynamics and Kinetics of Crambin 
Using an All-Atom Monte Carlo Simulation}} 

\author{Jun Shimada, Edo L. Kussell, Eugene I. Shakhnovich$^{\star}$}

\date{\today}
\maketitle
\vspace{0.5in}
\begin{center}
\textbf{Running title:} All-atom Monte Carlo Folding of Crambin \\
\vspace{0.5in}
\begin{spacing}{1}
Department of Chemistry and Chemical Biology \\
Harvard University \\
12 Oxford Street  \\
Cambridge MA 02138\\
\vspace{0.5in}
$^\star$corresponding author \\
tel: 617-495-4130 \\
fax: 617-495-3075 \\
email: eugene@belok.harvard.edu \\
\vspace{0.5in}
\end{spacing}
\end{center}

\newpage

\noindent
\begin{center}
\textbf{Abstract}
\end{center}
\noindent
We present a novel Monte Carlo simulation of protein folding, in which
all heavy atoms are represented as interacting hard spheres. This
model includes all degrees of freedom relevant to folding - all
sidechain and backbone torsions - and uses a G\={o} potential.  In
this study, we focus on the 46 residue $\alpha$/$\beta$ protein
crambin and two of its structural components, the helix and helix
hairpin.  For a wide range of temperatures, we have recorded multiple
folding events of these three structures from random coils to native
conformations that differ by less than 1 \AA\ dRMS from their crystal
structure coordinates.  The thermodynamics and kinetic mechanism of
the helix-coil transition obtained from our simulation shows excellent
agreement with currently available experimental and molecular dynamics
data.  Based on insights obtained from folding its smaller structural
components, a possible folding mechanism for crambin is proposed.  We
observe that the folding occurs via a cooperative, first order-like
process, and that many folding pathways to the native state exist. One
particular sequence of events constitutes a ``fast-folding'' pathway
where kinetic traps are avoided.  At very low temperatures, a kinetic
trap arising from the incorrect packing of sidechains was observed.
These results demonstrate that folding to the native state can be
observed in a reasonable amount of time on desktop computers even when
an all-atom representation is used, provided the energetics
sufficiently stabilize the native state.

\newpage

\nocite{TitlesOn}
\noindent
\large
\textbf{Introduction}\\
\normalsize

Previous lattice and off-lattice folding simulations featured
coarse-grained protein representations where amino acids are often
modeled as spheres
\cite{SSK2,EIS_COSB,PANDE_COSB,DILL_REVIEW,BRYN_REVIEW,KT,THIRUMALAI,BERRIZ,ZHOU,ONUCHIC}.
While such simplified representations have their advantages, such as
the reduction in the number of degrees of freedom, there are several
shortcomings.  First, one could argue that such models do not capture
the full complexity of conformational space, thereby not truly
addressing the Levinthal problem.  Second, coarse-grained models may
lack some realistic features of secondary structure elements.
Finally, such models do not address the packing of sidechains in the
protein interior, which is often viewed as an important aspect of the
folding process, given the diversity of sidechain shapes and their
dense packing in the native state ~\cite{PONDER}.

Molecular dynamics approaches have attempted to bridge the gap between
simulation and reality by using atomic representations of proteins
with explicit solvent molecules.  Unfortunately, the computational
times for such minimally coarse-grained simulations still remains
prohibitively high as evidenced by a recent work that yielded only a
partial folding trajectory~\cite{KOLLMAN} even with the aid of a
massively parallel supercomputer.  An ensemble of complete folding
trajectories is required in order to gain meaningful physical
insights, especially as our theoretical understanding of protein
folding has become increasingly grounded on statistical mechanical
principles \cite{EIS_COSB,ONUCHIC_WOLYNES,PANDE_RMP}.  Other molecular
dynamics research efforts have tried to address this issue by
obtaining multiple unfolding runs~\cite{DAGGETT,LAZARIDIS} at
extremely high temperatures.  Unfortunately, it is unclear whether the
observed unfolding pathways are indicative of folding pathways normal
conditions \cite{FINKELSTEIN_UNFOLDING}.

We present a Monte Carlo (MC) simulation~\cite{BINDER} which combines
an all-atom description of the protein with coarse-grained motions and
energetics.  In this simulation, (1) all heavy atoms in the protein
are represented as impenetrable spheres, (2) all backbone and
sidechain torsions, which account for all degrees of freedom relevant
to folding, are allowed to move, and (3) a square well, G\={o}
potential is used for the interaction energy.  There are several
advantages to this method.  First, a statistically significant number
of complete folding trajectories can be collected using conventional
computational resources.  Second, the atomic-level resolution of the
simulation yields detailed descriptions of the folding process of
actual protein structures including sidechain packing.  Finally, this
coarse-grained approach allows for a systematic investigation of the
physical principles dictating protein folding.  If one begins with a
detailed energy function as in molecular dynamics, it may be difficult
to deconvolute exactly which energy terms were essential (or
inessential) for folding.  On the other hand, as was demonstrated in
theoretical investigations~\cite{GROSBERG}, by thoroughly
investigating the successes and limitations of coarse-grained models,
it is possible to test which features of a model are necessary and/or
sufficient for describing the complex physics of heteropolymers.

\begin{figure}[tbp]
\begin{center}
\epsfig{file=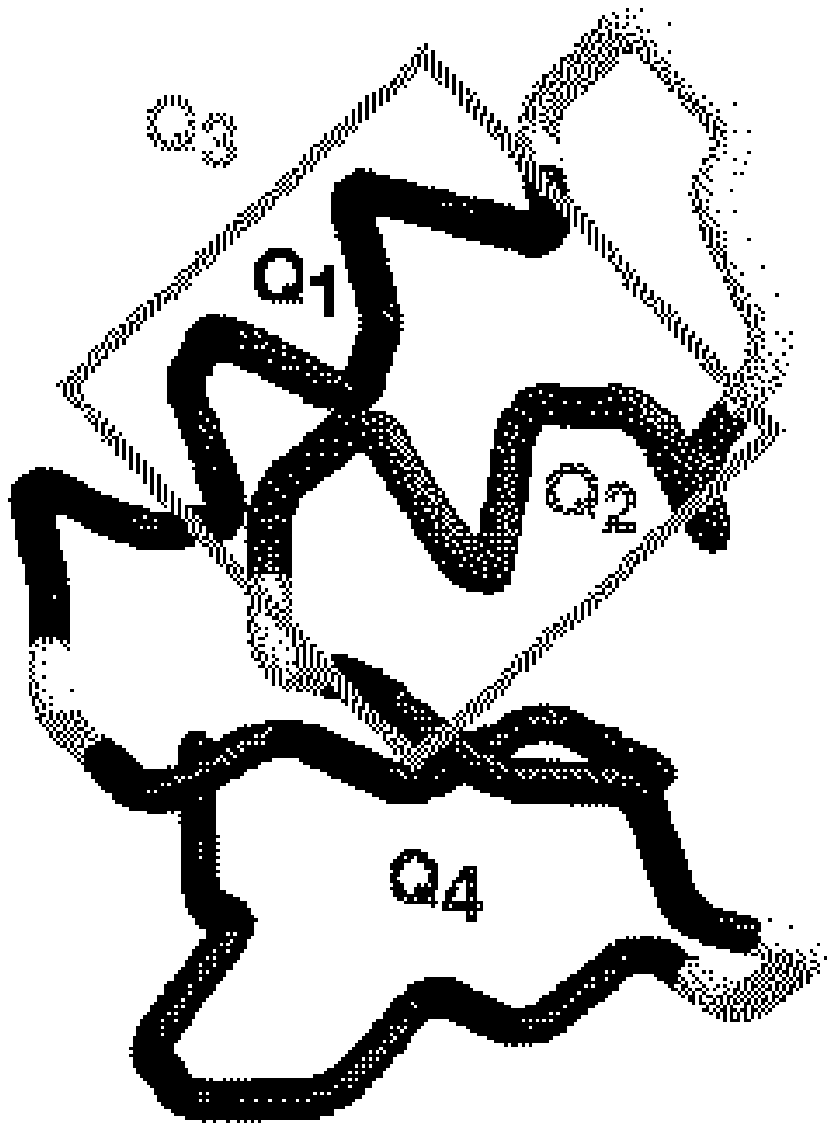}
\end{center}
\caption{}
\label{fig:NATIVE_CRAMBIN}
\end{figure}

Using our simulation, we have repeatedly folded the 46-residue
$\alpha-\beta$ protein crambin (Figure~\ref{fig:NATIVE_CRAMBIN}) to
within 1 \AA\ backbone dRMS (computed over all backbone atoms).
Furthermore, the thermodynamic and kinetic folding properties obtained
from our simulation are consistent with experimental studies of other
single domain proteins~\cite{BAKER,FERSHT_CI2}, for which the folding
transition generally exhibits two-state behavior with no accumulating
intermediates between the denatured and native states~\cite{JACKSON}.

One of the more successful methods reported in the literature
generated folded structures for crambin to $\approx$ 3 \AA\ C-$\alpha$
dRMS \cite{kolinski_skolnick} using a sequence-based potential.  These
final structures were obtained from a two-step heuristic approach:
simplified structures obtained from folding runs on a coarser lattice
were then refined on a finer lattice.  In contrast, our method uses a
potential based on knowledge of the native structure, but generates
the entire folding transition without altering key conditions (such as
the move set, temperature, or protein representation) once the
simulation is initiated.  As such, it produces an ensemble of
trajectories which can yield thermodynamic and kinetic information.
Although the potential we employ cannot be used to fold arbitrary
protein sequences, we demonstrate that the full conformational search
problem can be solved using standard computer resources.\\

\noindent
\large
\textbf{Theory and Motivation}\\
\normalsize

Analytical studies \cite{BRYNGELSON,RAMANATHAN,EIS_COSB,PANDE_RMP} and
lattice simulations \cite{EIS_AMG_PNAS,EIS_COSB} determined that only
certain sequences can fold in a biologically relevant time scale.
Such fast-folding sequences featured the native conformation as the
pronounced energy minimum.  For example, when evolution-like pressures
were applied to lattice protein sequences to preferentially select
mutations that improved folding speed (analogous to real-life
evolutionary pressures to select peptide sequences that survive
proteolysis), the resulting sequences featured a large energy gap
between their native and competing non-native states
\cite{AMG_VIA_EIS_PNAS}).  Conversely, in order to successfully
simulate the folding of biologically occurring protein sequences, it
is necessary to use a model that places the native state at the bottom
of the energy spectrum.

As noted in the Introduction, our folding model combines a
near-perfect geometrical representation of the peptide chain with a
potential energy function that delivers the required ``energy gap" by
explicitly assigning the native conformation as the ground state (see
Methods).  In general, a G\={o} potential is often viewed as providing
an idealized energy landscape by strongly biasing folding: first,
there is a strong correlation between energy and structural distance
(e.g. C-$\alpha$ RMS) from the native state; and second, the potential
energy surface is seen as very smooth and downhill when going from the
unfolded to the native state.

Despite this criticism, important aspects about folding kinetics can
be learned as demonstrated by several recent studies
\cite{ONUCHIC,PANDE_ROKHSAR,ZHOU}.  One particularly interesting
aspect is the role topology may play during the folding process
\cite{BAKER_NATURE_REVIEW}.  Since folding pathways are determined by
the barriers and local minima encountered on a free energy landscape,
non-trivial folding kinetics may be observed in a model that allows
for conformational entropy to compete with the G\={o} energy at the
relevant temperatures.  The predominance for frustration due to
topology can be problematic for long enough chains, but certainly
becomes pronounced when regular geometrical elements such as
$\alpha$-helices and $\beta$-sheets must be formed by bringing
together distant parts of the chain.  In fact, we demonstrate below
that if realistic protein geometry is used, severe kinetic traps due
to topological frustration make the folding kinetics extremely
complex.\\

\noindent
\textbf{The move set}\\

Our main concern when choosing the move set was to balance realism and
efficiency.  The natural motions of a solvated polymer chain at room
temperature occur via torsional degrees of freedom.  However, it
appeared that using only single torsional rotations may be
unrealistic, since it would lead to large conformational changes of
the polymer chain.  In solvent, large moves of such type are rare
because of the drag experienced by the moving chain.  A worse
situation is encountered in the compact state, as large rotations are
prohibited because of excluded volume restrictions.  To properly
address these issues, the torsional move set we chose has two
important features.  First, all of our moves consist of either two,
four or six concerted rotations.  This allows for a variety of
"loop"-like moves, which are particularly important when in the
compact state.  These rotations were required to be within 6 residues
of each other, in order to ensure that the majority of moves resulted
in local conformational changes.  Second, the step sizes for these
concerted rotations were drawn from a Gaussian distribution of
relatively small width.  More specificially, the probability for
selecting a backbone step size of $\theta$ was given by
\[ p(\theta) = \frac{1}{\sqrt{2 \pi \sigma^2}} \exp\left(-\frac{\theta^2}{2 \sigma^2}\right) \]
where $\sigma = 2^{\circ}$.  This ensures that a random walk is
performed in torsional space, which should likewise result in a random
walk in conformational space.  The vast majority of step sizes are
small enough that the conformational changes induced by our moves are
local.  Very rarely, when a fortutitous combination of torsions and
rotations are chosen and the chain is in an extended state, a global
conformational change is allowed.  During all of our simulations, such
global conformational changes were not observed in the compact state:
large conformational changes were always accompanied by partial
unfolding events.\\

\noindent
\textbf{Can our Monte Carlo simulation yield information about folding kinetics?}\\

MC simulations have been used in a wide range of problems in
statistical physics to obtain equilibrium information \cite{BINDER}.
As long as a MC simulation's move set satisfies the criterion of
detailed balance, convergence to a canonical distribution is
guaranteed.  A much more controversial aspect of MC simulations has
been its use in determining the relaxation kinetics to an equilibrium
distribution.  Because there is no explicit correspondence with real
time in a MC simulation, one might raise the objection
that all kinetic interpretations of MC trajectories are dependent on
the particular move set chosen.

For simple MC lattice folding simulations utilizing the standard
Verdier move set (consisting of the crankshaft, corner and tail flips)
\cite{VERDIER}, it was demonstrated that general conclusions on
folding kinetics, such as the importance of specific nucleation for
fast-folding, could be drawn if the analysis was performed on an
ensemble of folding runs \cite{ABKEVICH,VIA_BIOCHEMISTRY}.  This
approach was supported by an eariler study, which analytically showed
that the coil-globule relaxation time, as computed by the use of the
Verdier and other local move sets, matched theoretical
predictions\cite{HILHORST}.  Unfortunately, such rigorous analyses are
extremely difficult to complete as the complexity of the system being
simulated increases.

To justify our use of MC to interpret folding kinetics, we provide the
following heuristic argument.  Let $\vec{W}(s) = \{w_{i}(s)\}$
represent the probability distribution of states at step $s$ of a MC
simulation.  We can then represent the progression in our MC
simulation as the Markov process
\[ \vec{W}(s+1) = P \cdot \vec{W}(s) \]
where $P$ is the transition matrix, whose $i-j$ th element is given by
the probability to go from state $i$ to $j$.  In a Metropolis MC
simulation, the individual elements of this transition matrix obey
detailed balance, which means it is the equilibrium solution to the
series of first-order kinetic equations (also known as the master
equations)
\[ \frac{dw_{i}}{dt} = \sum_{j} -P_{ij}w_{i}+P_{ji}w_{j} = 0\].
Now consider a move set that is sufficiently local and unbiased
towards the native state.  During the relaxation to equilibrium, the
ensemble population will be highest where there are free energy
minima.  The evolution of the MC simulation thus leads to a gradual
shifting of the ensemble population from local minima to the global
minimum, the native state.  If kinetic events are separated by a large
enough number of local moves, and they are observed in an ensemble of
relaxation trajectories, they represent significant state population
shifts and reflect properties of the free energy landscape (e.g., the
depth of local minima and barrier heights).  In this manner, examining
an ensemble of trajectories highlights the major kinetic events during
folding by averaging out short MC time events.  Importantly, this has
the effect of minimizing differences arising from the selection of a
particular move set.

The kinetic picture arising from a MC simulation is thus
coarse-grained.  If a common free energy landscape is used, both MC
and other types of simulations should agree on the sequence of major
folding events, if they are sufficiently separated in time.  This was
demonstrated by Rey and Skolnick (1991), who concluded that a MC
simulation yielded the same folding pathways for an $\alpha$-helical
hairpin as a Brownian dynamics simulation, for which there was real
time evolution.  We emphasize that MC simulations are not immediately
justified in making quantitative predictions or microscopic analyses
of folding kinetics.  In order to reliably predict folding rates,
every MC simulation needs to be extensively calibrated to existing
experimental data.  It is likely that such calibration will rule out
several move sets.

Finally, perhaps the most important requirement of achieving
quantitative agreement with experimental kinetic data is that the free
energy landscape must be sufficiently realistic.  Even molecular and
Langevin dynamics simulations, where a well-defined time step is used
to evolve the simulation, will be unable to make quantitative -- and
perhaps even qualitative -- analyses of folding kinetics if this
requirement is not met.  Because there have been no reports of a
successful \emph{ab initio} folding potential, we decided that the G\={o}
potential, combined with an all-atom representation of the protein,
was the best choice for this simulation.

Before proceeding to analyze the folding kinetics of crambin, we
decided to more accurately assess the validity of our move set by
examining the thermodynamics and kinetics of the helix-coil
transition, for which there are experimental and molecular dynamics
data.\\

\noindent
\large
\textbf{Results}\\
\normalsize

The formation of the $(i,i+4)$ hydrogen bonding scheme in helical
structures results in a high native contact density.  As a result, a
nondiscriminate use of G\={o} potentials will lead to unusually strong
energetic biases towards helices.  We thus started by calibrating the
stability of the two crambin helices with the predictions of AGADIR
\cite{AGADIR} (at pH = 7 and $T = 298$ K): at temperature relevant for
our crambin folding simulations, both helices had zero helical
propensities.  Elimination of backbone contacts was the simplest way
to dramatically reduce helix stability without introducing biases due
to sequence identity.  All subsequent data was collected while using
this slightly modified G\={o} potential.\\

\noindent
\textbf{Helix-coil transition}\\

\begin{figure}[tbp]
\begin{center}
\epsfig{file=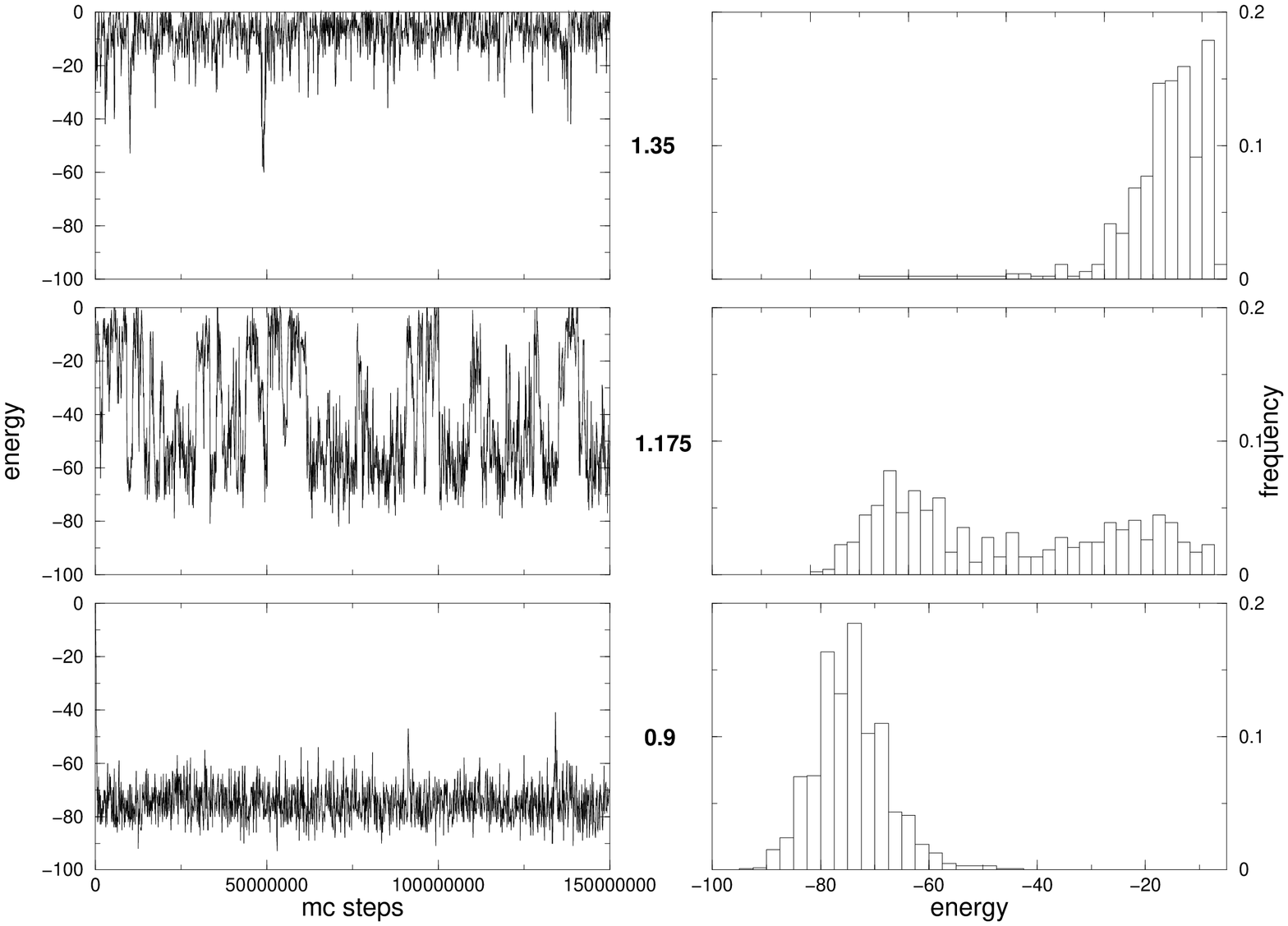,angle=-90,width=6in}
\end{center}
\caption{}
\label{fig:HELIX_RUNS}
\end{figure}

\begin{figure}[tbp]
\begin{center}
\epsfig{file=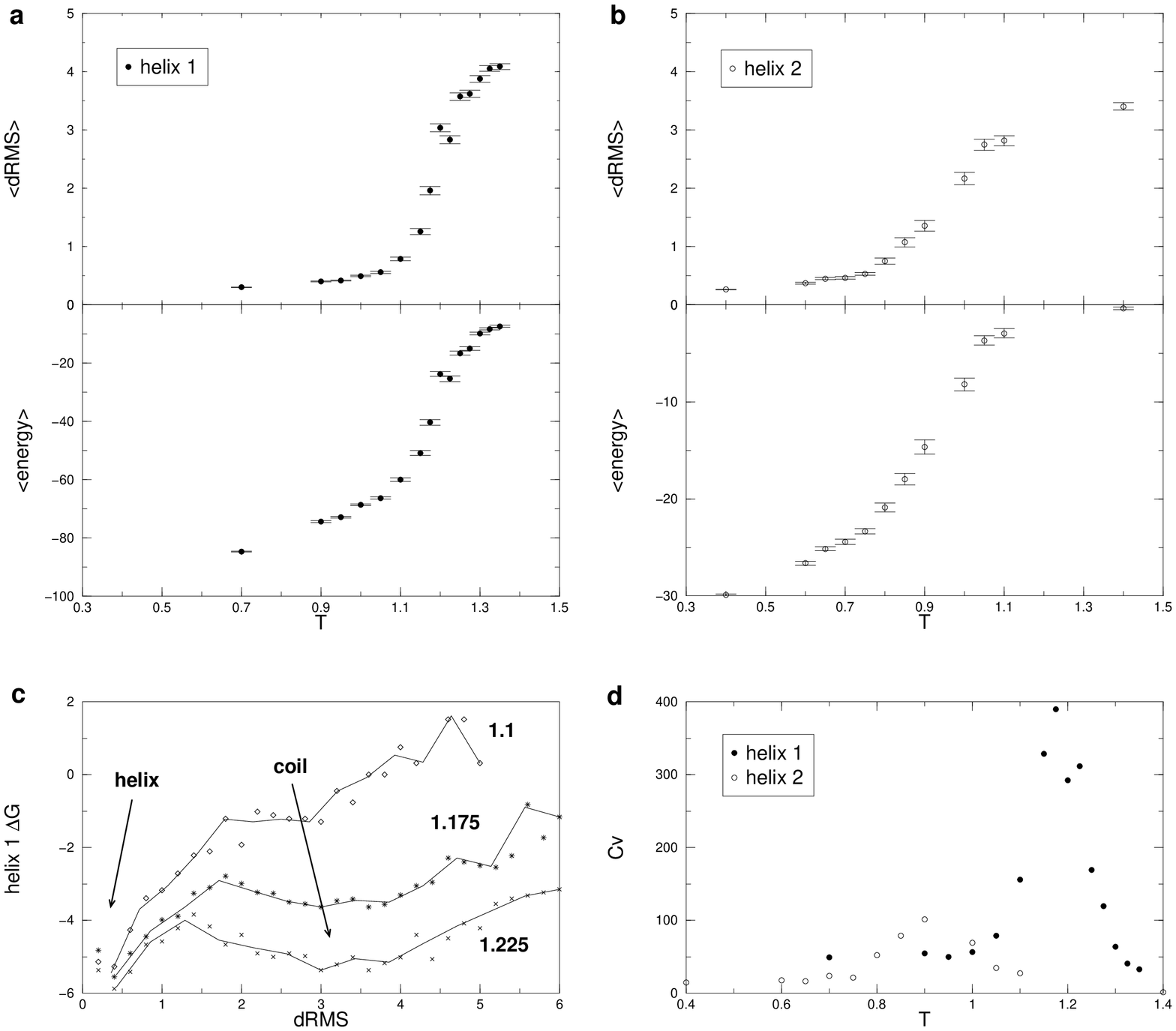,angle=-90,width=6in}
\end{center}
\caption{}
\label{fig:HELIX_THERMO}
\end{figure}

Both helix 1 and 2 (see Figure \ref{fig:NATIVE_CRAMBIN} for sequences)
exhibited an abrupt transition to the coil state, with the transition
temperatures occurring at $T_{f}^{\textrm{helix 1}} \approx 1.2$ and
$T_{f}^{\textrm{helix 2}} \approx 0.9,$ respectively (Figures
\ref{fig:HELIX_THERMO}a \& b).  The energy histograms show a sudden
shift in helix and coil state populations as $T_{f}^{\textrm{helix
1}}$ is passed, with a broadening of the distribution at
$T_{f}^{\textrm{helix 1}}$ (Figure \ref{fig:HELIX_RUNS}).  The free
energy curves clearly demonstrate the emergence of a local, second
minimum centered at 3.5\ \AA\ backbone dRMS for $1.1 < T <
T_{f}^{\textrm{helix 1}}$ (Figure \ref{fig:HELIX_THERMO}c).  At
$T_{f}^{\textrm{helix 1}}$, this second minimum becomes equal in free
energy to the helix state minimum, resulting in an equilibrium of both
helix and coiled states.  Finally, the heat capacities for both
helices are sharply peaked around $T_{f}$ (Figure
\ref{fig:HELIX_THERMO}d).  Based on the width of the heat capacity
peak at $T_{f}$, it appears that the helix 2 shows a weaker
transition, which is expected for a shorter helix.  These
thermodynamic observations fully agree with Zimm-Bragg theory
\cite{ZIMM_BRAGG}, simulations \cite{HUMMER,DAURA1,DAGGETT_HELIX}, and
experiment \cite{THOMPSON1}.

\begin{figure}[tbp]
\begin{center}
\epsfig{file=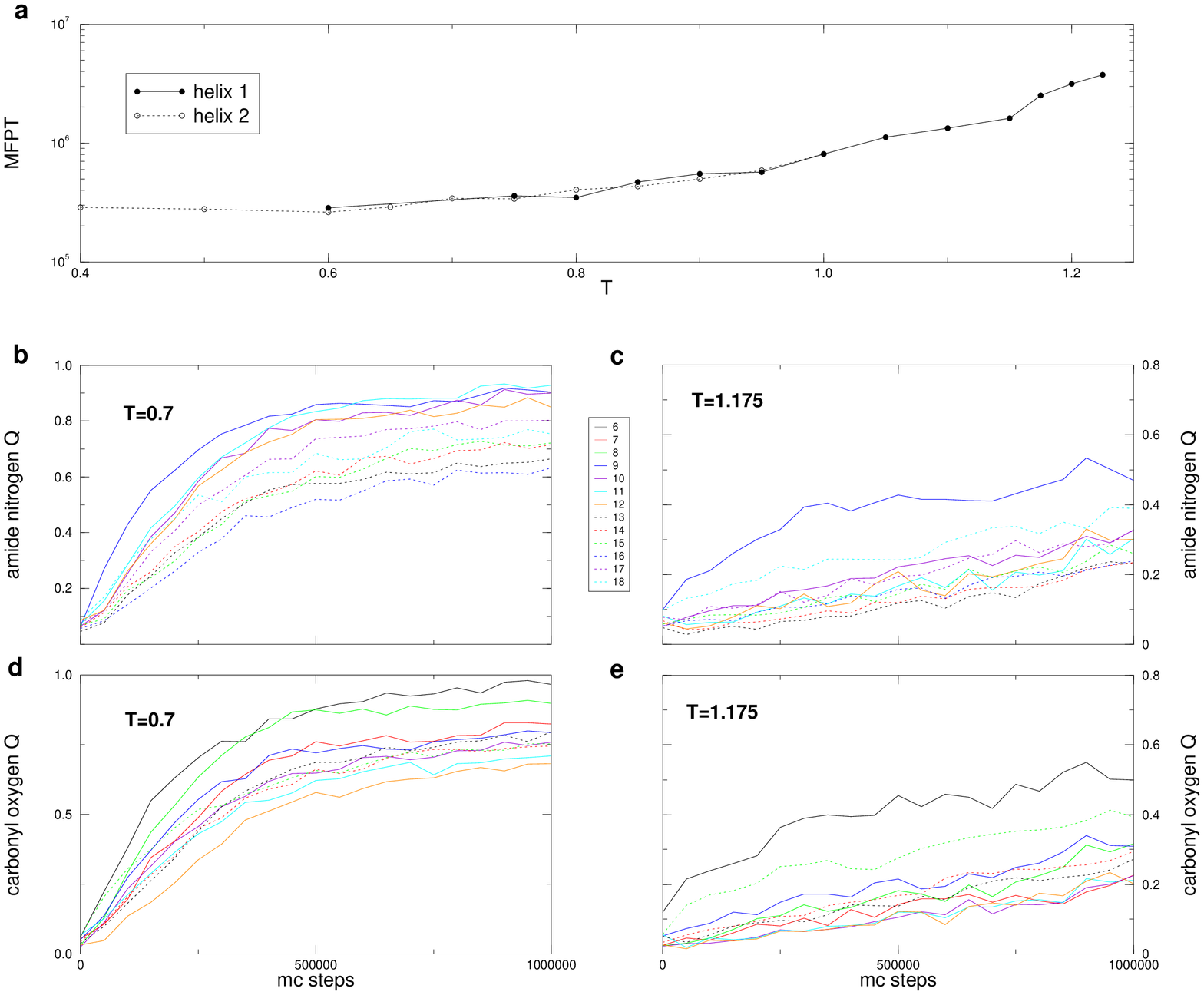,angle=-90,width=6in}
\end{center}
\caption{}
\label{fig:HELIX_KINETICS}
\end{figure}

The median first passage time (FPT) for helix 1 rose rapidly as T was
increased for $T > 1.1$ (Figure \ref{fig:HELIX_KINETICS}a).  Given
that the G\={o} energy more or less increases with increasing dRMS, it
is clear that the free energy barrier to helix formation (Figure
\ref{fig:HELIX_THERMO}c) is purely entropic.  Interestingly, for $T <
1.1$, the helix 1 formation rate becomes weakly dependent on
temperature, suggesting that there are no major free energy barriers.
At all temperatures, the distribution of FPTs exhibited long tails,
but more data was needed to accurately determine whether the
distribution was non-exponential.  As noted by \cite{HUMMER,HUMMER2},
the formation rate appears to be dominated by a diffusive search for a
nucleation event.  Once the nucleus is formed, the formation of the
rest of the helix is rapid (see the trajectory at $T \approx T_{f}$ in
Figure \ref{fig:HELIX_RUNS}).

Examination of the ensemble helix 1 formation kinetics shows that ALA
9 is the dominant nucleation site at both low and high temperatures
(Figures \ref{fig:HELIX_KINETICS}b-e).  This is shown by the rise in
the fraction of native contacts (Q) made by the amide nitrogen of ALA
9.  It is necessarily accompanied by rise in the carbonyl oxygen Q of
SER 6, as ALA 9 and SER 6 make key sidechain-sidechain contacts to
initiate the formation of the first turn.  A logical explanation for
this dominant nucleation event is the absence of sidechain entropy at
ALA 9 in our model, thus lowering the free energy cost of constraining
this residue in a helical conformation.  Upon nucleation, at higher
temperatures, the propagation appears to rapidly proceed towards the
N-terminus (Figures \ref{fig:HELIX_KINETICS}d \& e), while at lower
temperatures, this asymmetry seems to be weaker (Figures
\ref{fig:HELIX_KINETICS}b \& c).  This may be explained by the larger
entropic cost of constraining the large sidechain of ARG 10 at higher
temperatures.  A similar observation was made by Pande \emph{et al.}
(personal communication, 2000), whose simulation showed that helix
propagation was slowed by the presence of arginine residues in a
poly-alanine helix.  The two exponential relaxations processes
detected in the A$_{8}$-R-A$_{4}$ peptide by Thompson \emph{et al.}
(1997) may be explained by a similar phenomenon: the fast quenching of
the N-terminus fluorescent label is due to the rapid
nucleation/propagation of the alanine rich regions (from both N- and
C-termini) and the slow phase due to the slow incorporation of the
arginine residue into the growing helix.

At lower temperatures, SER 11 appears as a competing nucleation site.
Like ALA 9, SER 11 is a low entropy residue.  However, in order for
SER 11 to make helical contacts with neighboring residues, it requires
constraining at least one of three large sidechains (ARG 10, ASN 12,
PHE 13), which may be unfavorable for entropic or steric reasons.  In
contrast, forming the SER 6 - ALA 9 interaction requires constraining
smaller residues.  For this reason, it seems to be more favorable to
nucleate the helix at ALA 9.  In support of this idea, we note that
SER 11 disappears as a nucleation site at higher temperatures.

A third nucleation site occurs at LEU 18.  This appears to be driven
primarily by energy, as many native sidechain-sidechain contacts are
made with VAL 15.  However, as the effect of sidechain entropy is
increased as temperature is raised, we see that the VAL 15 - LEU 18
interaction forms slower relative to the SER 6 - ALA 9 interaction
than at lower temperatures.  Interestingly, raising the temperature
seems to have a larger effect in reducing the capacity of SER 11 to
nucleate the helix compared to LEU 18.

Although the propagation asymmetry predicted by Young \emph{et
al.} (1996) was not observed at all temperatures, the
philosophy behind their general approach seems to be correct:
preferences in helix propagation and nucleation are determined by the
specific atom-atom interactions.\\

\noindent
\textbf{Folding simulations of crambin}\\

Given that our move set yielded reasonable kinetic and thermodynamic
helix-coil transition data, we proceeded to fold crambin.  Starting
from random coils, we considered the protein folded if the following
three criteria were satisfied for at least $5 \times 10^{6}$ steps:
(1) the fraction of native contacts ($Q$) exceeded 0.7; (2) the
backbone dRMS was less than 1.25 \AA; (3) the fraction of native
contacts in the four secondary/tertiary structural features ($Q_{i}$,
$1 \leq i \leq 4$) exceeded 0.5 (see Figure~\ref{fig:NATIVE_CRAMBIN}).
Criterion (1), which is similar to the one used to signify a folding
event in lattice simulations~\cite{ABKEVICH}, could not by itself
distinguish conformations that one might intuitively consider folded
(i.e., properly formed secondary structure and low overall dRMS) from
obvious misfolds.  Although our definition for folding did not measure
when equilibrium was attained, it was useful for identifying when the
major folding event -- the transition from the random coil to a
near-native state -- occurred.\\

\noindent
\textbf{Thermodynamics}\\

\begin{figure}[tbp]
\begin{center}
\epsfig{file=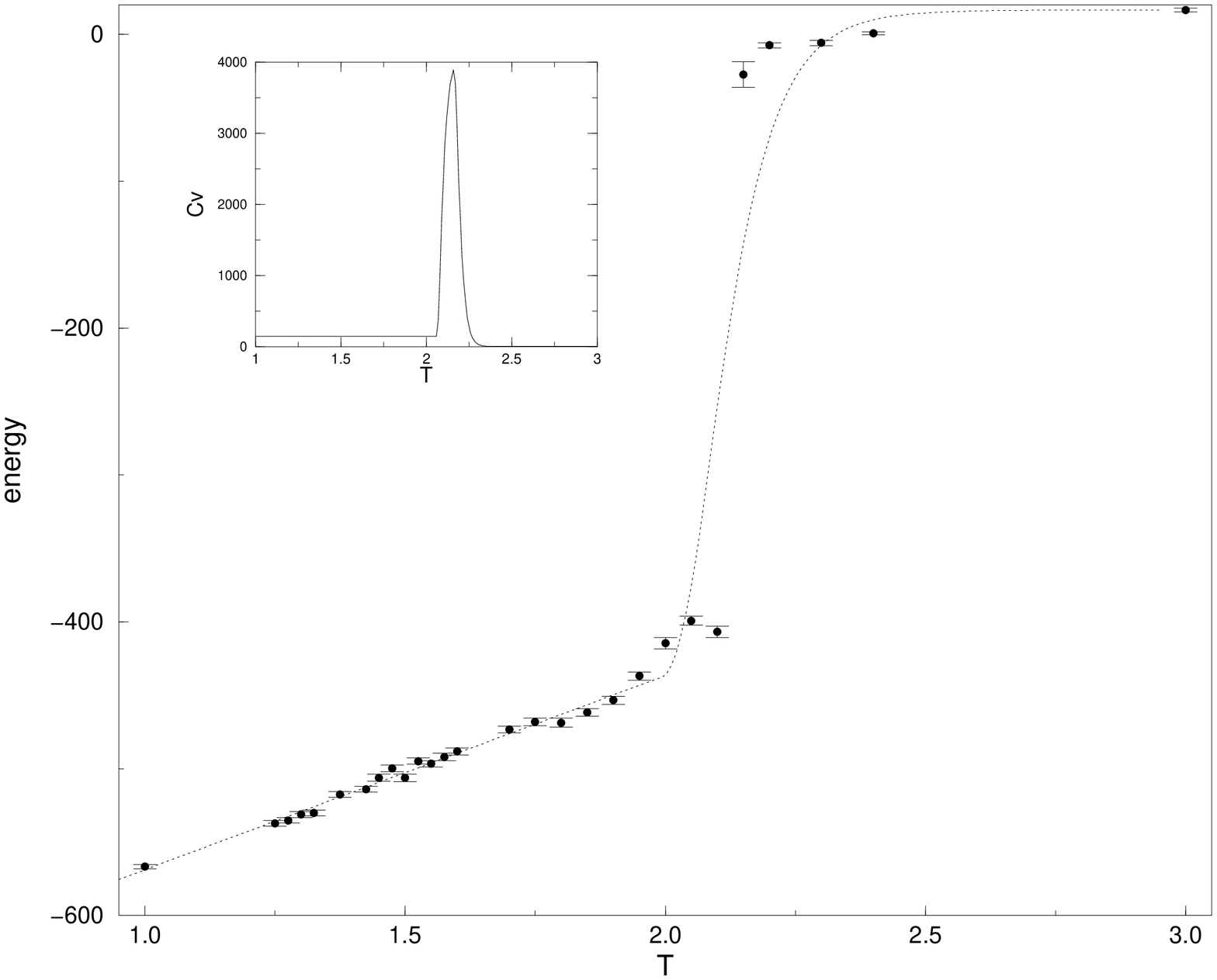,angle=-90,width=7in}
\end{center}
\caption{}
\label{fig:STABILITY}
\end{figure}

The folding thermodynamics is aptly
described by a first order-like transition~\cite{HAO}
(Figure~\ref{fig:STABILITY}), with the MC folding transition
temperature ($T_{f}$) estimated to be between 2.1 and 2.3.  For high
temperatures ($2.0 < T < 3.0$), the data is well-fit by the
exponential function typically used to describe two-state folding
behavior~\cite{FERSHT_BOOK}.  This is in general agreement with the
folding thermodynamics obtained with lattice models \cite{EIS_COSB}.
By matching the backbone dRMS value from NMR measurements (1.48 \AA)
\cite{NMR_CRAMBIN}, we estimate that room temperature corresponds to
$2.1 < T < 2.2$, which is consistent with the general observation that
most proteins are marginally stable at room temperature
\cite{CREIGHTON}.\\

\noindent
\textbf{Kinetics}\\

\begin{figure}[tbp]
\begin{center}
\epsfig{file=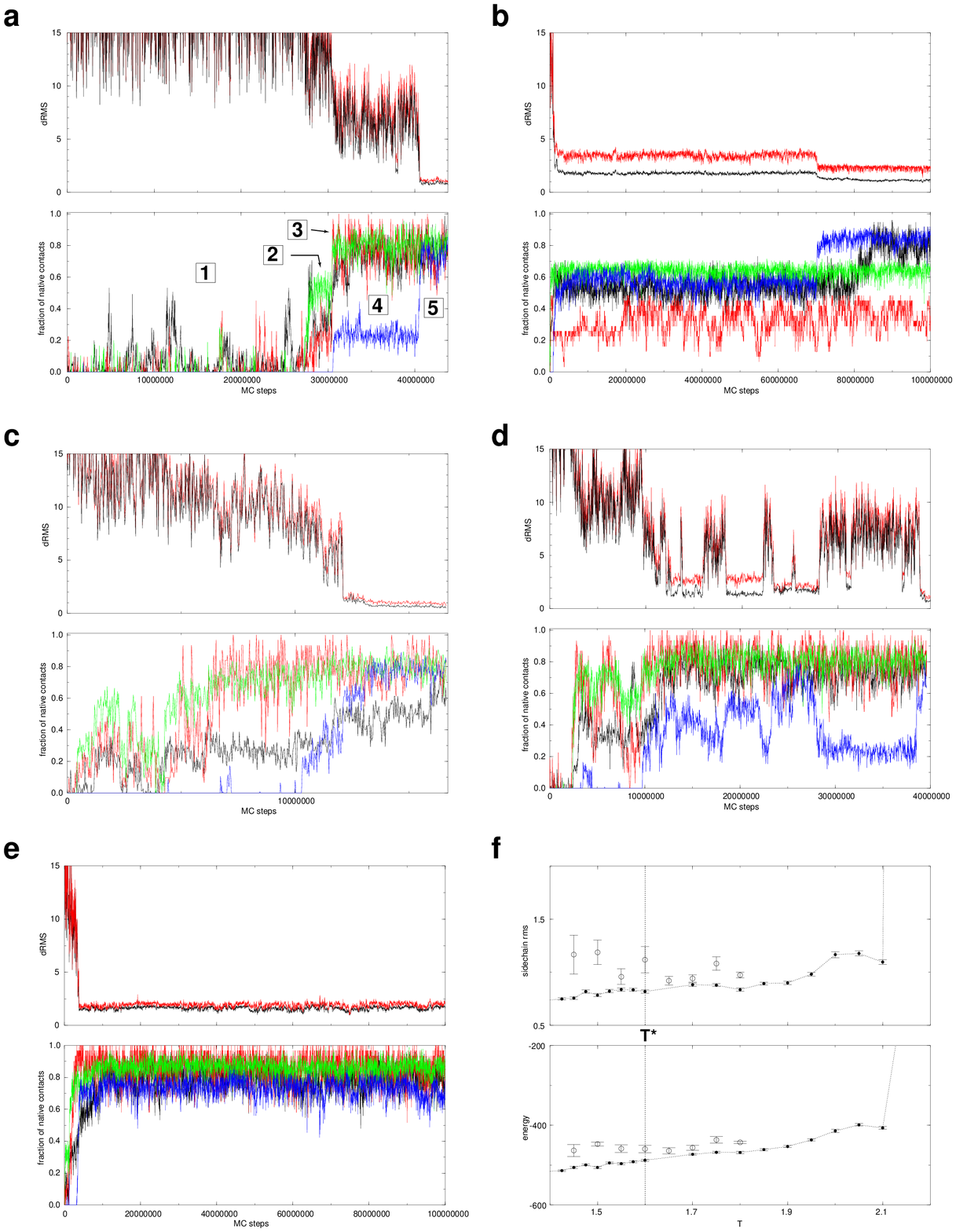,height=7in}
\end{center}
\caption{}
\label{fig:RUNS}
\end{figure}

\begin{figure}[tbp]
\begin{center}
\epsfig{file=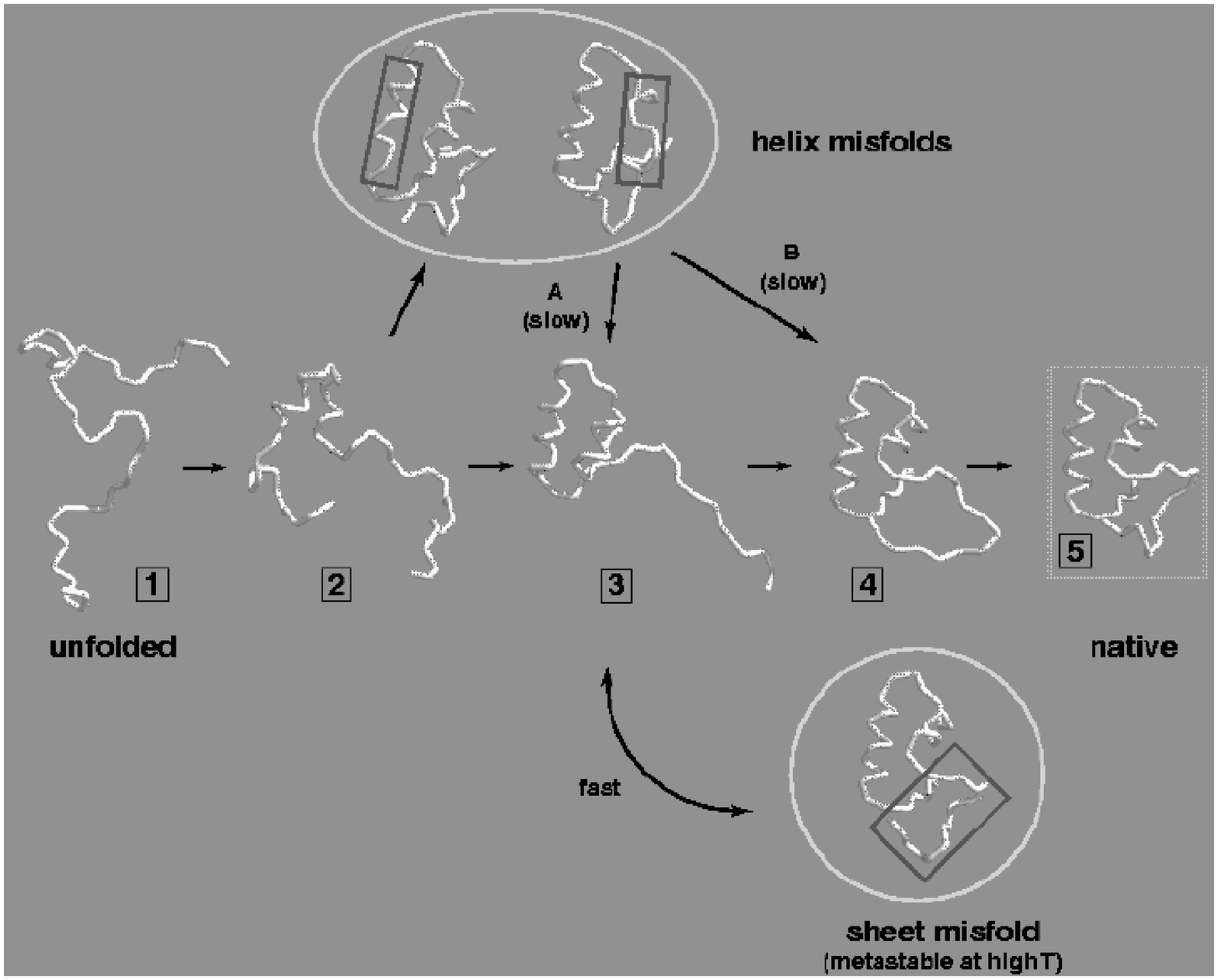,angle=-90,width=7in}
\end{center}
\caption{} \label{fig:PATHWAY}
\end{figure}

Representative folding runs are shown in Figures
\ref{fig:RUNS}a-e.  Although different folding pathways were observed
(Figure \ref{fig:PATHWAY}), a fast-folding pathway was found where
kinetic traps were avoided.  The ensemble of trajectories following
this fast-folding pathway was characterized by a definite sequence of
events: (1) formation of the inter-helix contacts (event 2 in Figure
\ref{fig:PATHWAY}); (2) formation of the two helices (event 3); and
(3) formation of the $\beta$-sheet (events 4 and 5).  This observation
can be rationalized from simple topological considerations: the
helices are most easily formed when the two ends of the polymer are
not constrained by the $\beta$-sheet.

\begin{figure}[tbp]
\begin{center}
\epsfig{file=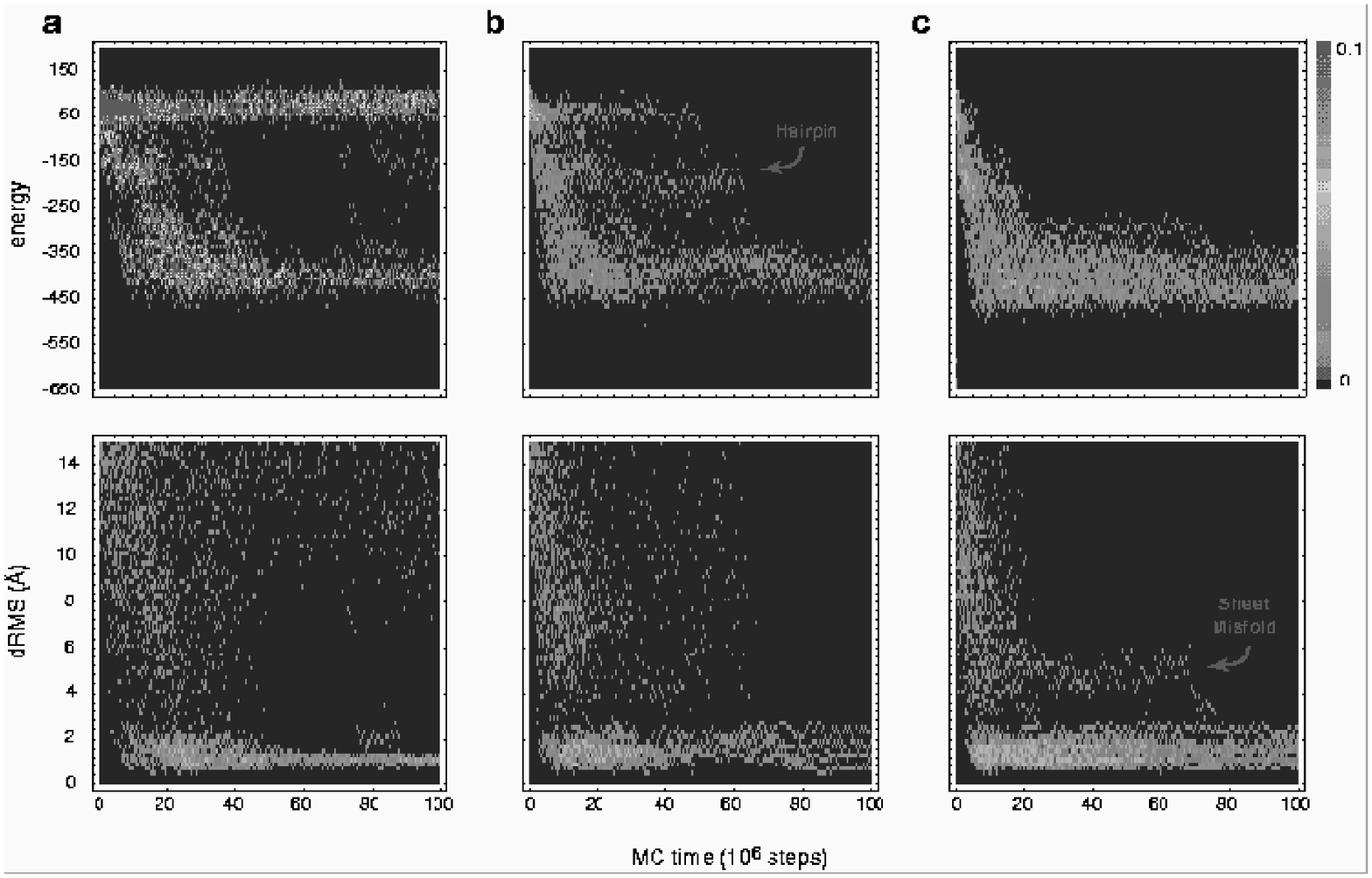,angle=-90,width=7in}
\end{center}
\caption{} \label{fig:KINETICS}
\end{figure}

At high temperatures ($1.7 < T < 1.8$), at least half of the folding
trajectories did not collapse within $10^8$ steps, making clear that
collapse is the rate-limiting step (Figures \ref{fig:RUNS}a,
\ref{fig:KINETICS}a).  The kinetics appear to be two state, as
evidenced by the narrowly distributed near-native and unfolded
populations, rapid collapse events, and no accumulating intermediates.
The transiently populated state with energy -$150$ corresponds to
event 2 of the fast-folding pathway (Figure ~\ref{fig:PATHWAY}) where
a subset of the inter-helix and helix contacts have been formed.  At
middle temperatures ($1.55 < T < 1.675$), this intermediate
accumulates for longer times and the collapsed state ensemble broadens
(Figure ~\ref{fig:KINETICS}b).  At low temperatures ($T < 1.525$) this
effect becomes particularly pronounced, with high energy, collapsed
states accumulating for very long times (Figure ~\ref{fig:KINETICS}c).
While collapse is very fast at these temperatures, the persistently
broad distribution of low dRMS ($< 2$ \AA) states indicates that the
compact state is riddled with deep traps.  Finally, the median folding
time rapidly increased for $T < 1.4$ and $T > 1.8$, and a broad
minimum existed at $1.5 < T < 1.65 $.  More runs are needed
to narrow the fastest folding temperature to a smaller range.\\

\noindent
\textbf{Role of helix stability in crambin folding}\\

The zero helix stability at crambin folding temperatures likely
explains why the diffusion-collision scenario \cite{WEAVER} was not
observed as the major fast-folding pathway; in fact, only 1 out of
over 150 runs which folded had the helices forming prior to the
inter-helix contacts.  It is emphasized that the
present move set is not likely to contain any biases against
diffusion-collision kinetics.  Unlike lattice MC simulations, in which
helices can reorient only by partial unfolding, our model allows
entire secondary structure elements to move as units.  To confirm this
point, we performed folding simulations of the two helix hairpin motif
(residues 6-30) in isolation.  

\begin{figure}[tbp]
\begin{center}
\epsfig{file=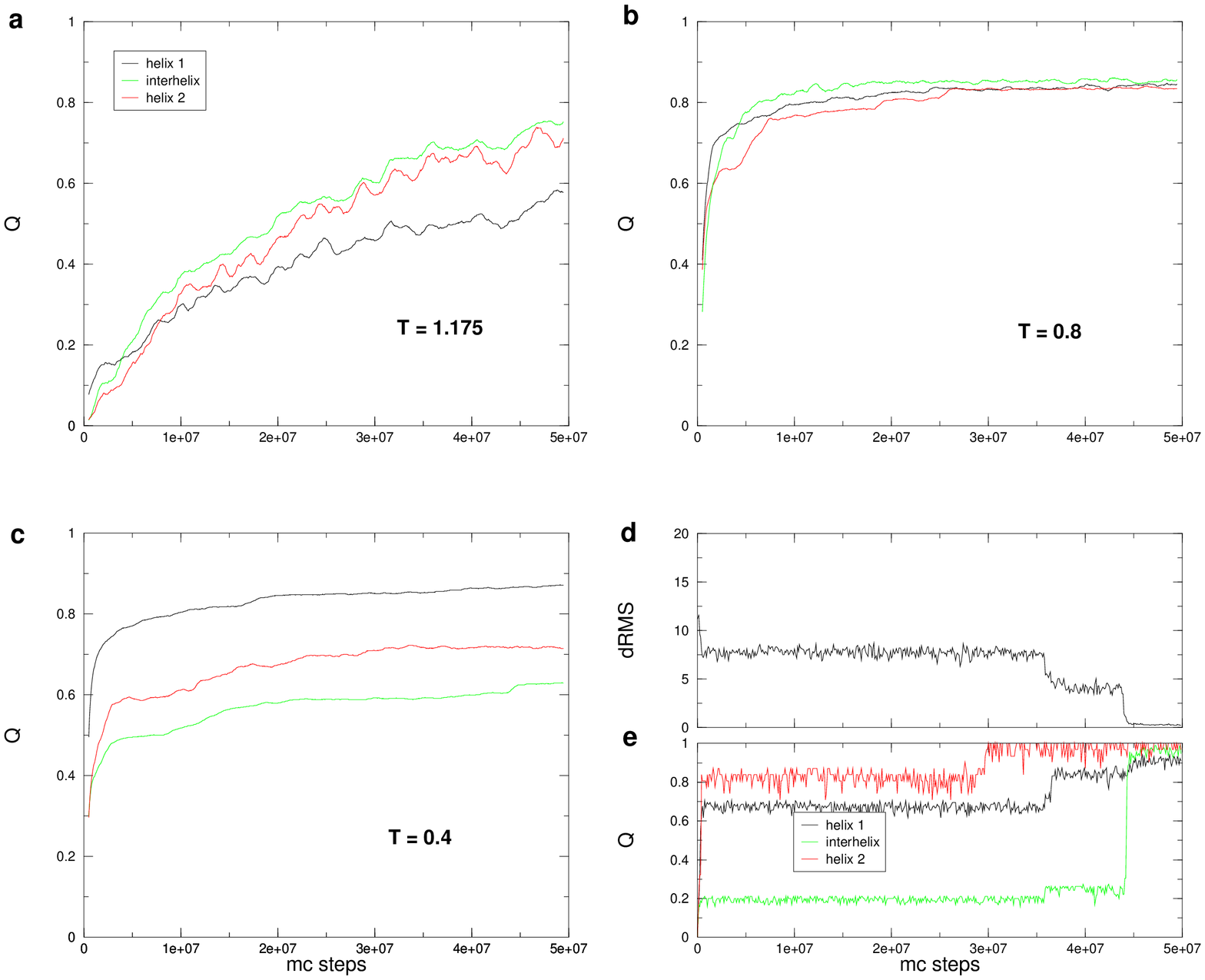,angle=-90,width=7in}
\end{center}
\caption{} \label{fig:HAIRPIN}
\end{figure}

In general, if the temperature is sufficiently lowered, we readily
observed diffusion-collision kinetics.  A representative
diffusion-collision folding trajectory of the hairpin at T=0.4 is
shown in Figure \ref{fig:HAIRPIN}d.  From the ensemble data, it
is clear that the diffusion-collision scenario emerges as the dominant
kinetic pathway if helix formation is rapid and the helical state is
stable (Figures \ref{fig:HAIRPIN}a-c).  This requires that the
temperature be sufficiently low ($T < 0.8$).  Since marginally stable
$\beta-$hairpin formation is inherently slow \cite{FINKELSTEIN_BETA},
at low temperatures where helices fold fast, there is a clear
separation of time scales: the helices form first, followed by the
hairpin.  Given that folding occurs at temperatures where helices are
unstable, we must thus rule out diffusion-collision as an important
kinetic pathway for this protein

Using a C-$\alpha$ off-lattice model with a G\={o} potential, Zhou
\emph{et al.} observed that for a three-helix bundle the "fast track"
folding pathway, where no kinetic intermediates are encountered, was
the diffusion-collision pathway.  In light of our hairpin data, this
result is not surprising.  The three-helix bundle was being folded at
extremely low temperatures: for a bias gap of 1.3 (native interaction
= -1; non-native interactions = +0.3) , the protein was being folded
at a temperature that was $\approx $ 25\% of the collapse transition
temperature ($T_{f}$).  On our scale, we began to observe crossover to
diffusion-collision behavior at $0.8/2.2 \approx 0.36 T_{f}$.
Although the parameters of the G\={o} potential are slightly
different, it is likely that Zhou \emph{et al.} were working at
extremely low temperatures where both the helices and the protein were
very stable.  If we estimate $T_{f}$ for actual proteins to be roughly
350 K, this implies that the folding simulations by Zhou \emph{et al.}
were completed at less than 100 K.  It is also known experimentally
that only the third helix of the bundle is marginally stable ($\approx
30\%$ helical content) under normal folding conditions \cite{BAI}.
Folding a similar three-helix bundle using Langevin dynamics, Berriz
and Shakhnovich found that the diffusion-collision behavior was
observed at low temperatures but disappeared as the temperature was
raised (GB and ES, unpublished data).  For these reasons, it
will be interesting to see if diffusion-collision behavior continues
to be observed by Zhou \emph{et al.} at higher temperatures.

\noindent
\textbf{Characterization of traps}\\

The kinetic traps observed were either backbone misfolds, where the
secondary structure elements were not properly formed, or compact
conformations with incorrectly packed sidechains.  The first type of
backbone misfold resulted when the $\beta$-sheet folded incorrectly
after the helix-turn-helix moiety was formed.  Correction of this
misfold required partial or complete unfolding of the $\beta$-sheet.
At high temperatures ($1.7 < T$), this misfold was metastable, as the
$\beta$-sheet repeatedly folded and unfolded until the correct
topology and packing was achieved (Figure ~\ref{fig:RUNS}d).  In
general, forming the sheet was observed to be the rate-limiting step
at low temperatures ($T < 1.5$) (see Figure \ref{fig:KINETICS}c).  The
other type of backbone misfold occurred when the two helices were not
formed properly prior to the formation of the $\beta$-sheet.  Two
pathways (A and B in Figure \ref{fig:PATHWAY}) were available at all
temperature to correct these helix misfolds.  Pathway A required the
$\beta$-sheet to partially or completely unfold.  In contrast, pathway
B corrected the helix while keeping the sheet intact.  This pathway
was accelerated with increasing temperature, as the ``breathing
motion'' resulting from greater backbone fluctuations facilitated
reinsertion of the helix sidechains (compare, for example, the $Q_{4}$
fluctuations in Figures \ref{fig:RUNS}c and d).
 
The traps resulting from incorrectly packed sidechains (Figure
\ref{fig:RUNS}e) are characterized by low backbone dRMS and correct
backbone topology (high $Q_{i}$'s) and were observed only at low
temperatures.  At higher temperatures, after the major folding event,
equilibrium sidechain packing is rapidly achieved.  As shown in Figure
\ref{fig:RUNS}f, for $T < T^{*} \approx 1.6$, near-native
backbones (dRMS $< 1.25$ and $Q_{i}$'s $> 0.6$) obtained from folding
runs would not relax further to match the energies attained in runs
initiated from the native state.  This suggests that $T^{*}$ may
signify a kinetic transition temperature, where ergodicity is broken
and a gap emerges between measurements taken from finite but long
unfolding and folding runs.  At low temperatures, it appears that the
major backbone folding event traps sidechains in disordered non-native
conformations, which cannot be readily relaxed because of insufficient
backbone fluctuations.\\

\noindent
\large
\textbf{Discussion}\\
\normalsize

Compared to molecular dynamics studies of solvated folding
\cite{KOLLMAN}, the approach used in this study is still minimalist.
Yet, we have demonstrated that important insights into the folding
process, such as the role of sidechain packing, may be obtained by
properly combining an all-atom description with simple, atomic
resolution energetics.  Importantly, our simulation can record a
statistically significant number of folding events at atomic
resolution for real protein sequences, thereby allowing a relatively
direct comparision between simulation and experimental results.

Unfortunately, because of the lack of experimental folding studies on
crambin, our results presently cannot be directly verified. They
should be viewed as theoretical predictions which may be tested by
future experiments.  We selected crambin for this study because of its
small size and its non-trivial $\alpha-\beta$ structure.  We note that
as the temperature approaches $T_{f}$, the folding kinetics of crambin
approaches two state behavior, in agreement with experimental studies
of most small single domain proteins \cite{JACKSON}.  It is plausible
that under a G\={o} potential, where all native interactions are
treated identically, the folding properties of crambin should be
consistent with those of similar size and topology.  

Furthermore, the modern viewpoint that nucleation is a major event in
the folding kinetics \cite{FERSHT_PNAS,SOSNICK,VIA_BIOCHEMISTRY} is
consistent with our results: at temperatures just below $T_{f}$,
collapse via formation of the $\beta$-sheet is the major kinetic
barrier (Figure \ref{fig:RUNS}a).  A transiently populated, partially
collapsed intermediate in which the helix hairpin is formed is seen
(Figure \ref{fig:KINETICS}a), and is probably necessary for proper
sheet formation.  Once the sheet has formed, the chain is compact, and
folding proceeds rapidly with concomitant sidechain packing.  With the
individual helices unstable in isolation at folding temperatures, the
hairpin formation did not follow a diffusion-collision pathway.
Rather, since the necessary event for hairpin formation is the
localization of inter-helix contacts, the kinetic mechanism is also
likely to be a nucleation event, similar to the one theorized for
$\beta$-hairpins \cite{FINKELSTEIN_BETA}.

In addition, structural elements that bring together residues that are
closer in position along the chain (such as the helix hairpin)
appear to form faster than those which bring together residues distant
in sequence position.  We believe this is similar to the observation
made by Plaxco \emph{et al.} that folding kinetic rates are correlated
with the relative contact order of the native state structure
\cite{PLAXCO}.

Even with a potential strongly biased towards proper folding, the
presence of diverse sidechain geometries and excluded volume
interactions can lead to the presence of severe kinetic traps, as we
observe at very low temperatures.  However, our results demonstrate
that at reasonable temperatures sidechains can be successfully packed
in a manner consistent with a low dRMS native-like backbone
conformation.  We believe that the move set we employed contributed to
the success of our simulation.  Efficient sampling in the compact
state, as evidenced by the $>10\%$ acceptance rate even for very low
temperatures, resulted from both global and local conformational
changes being permitted, with global changes becoming more
available with higher temperatures.  The qualitative folding behaviour
is consistent with experimental observations of small single domain
proteins, suggesting that the essential features of a polypeptide with
all torsional degrees of freedom are captured by this move set.

It is important to note that the helix-coil transition kinetics
obtained from our simple model qualitatively showed agreement with
experiment and molecular dynamics simulations.  Furthermore, whereas
current state-of-the-art molecular dynamics simulations focus on
extremely short helical segments (penta- and heptapeptides) to study
the the helix-coil transition \cite{HUMMER2,DAURA1}, we have been able
to carry out statistical mechanical analyses on full length helices.
The kinetics of early events during folding, such as helix formation,
which occur on the timescale of hundreds of nanoseconds, can thus be
investigated with our model provided the folding mechanism is examined
from an ensemble viewpoint.  Since the MFPT for helix formation is
roughly 1 million MC steps, this puts the folding of crambin (50-100
million MC steps) in the microsecond range, which is reasonable for a
protein of its size.  Recent data suggests that the protein G
$\beta-$hairpin takes 50 million MC steps to fold (JS, ELK, EIS,
unpublished data), which places it in the microsecond range to fold in
real time, nearly matching the $\approx 10$ microsecond rate observed
experimentally \cite{MUNOZ_THOMPSON}.  Furthermore, early data on
protein G folding (JS, ELK, EIS, unpublished data) indicates a folding
rate of a few billion MC steps.  This is one order of magnitude faster
than the experimental rate (a few milliseconds, \cite{BAKER_NSB}) but
the qualitative agreement is promising.  It particularly encouraging
to note that the folding rates of secondary structure elements as
observed in our simulation are properly separated in timescales.  This
justifies the use of our simulation to draw qualitative conclusions
about kinetic events involving the formation and/or organization of
secondary structure elements.

We intend to carry out similar studies on larger single domain
proteins for which there are extensive experimental data.  The faster
folding times we have preliminarily observed for protein G is
undoubtedly because of the G\={o} landscape, which presents an
idealistic, perfectly downhill energy landscape.  With the development
of a sequence-based potential for our model (currently in progress),
we expect our correspondence with experimental rates will improve.
Taking our results as a proof-of-concept, we believe that detailed
investigations of folding pathways may finally be possible using
ordinary computational resources.\\

\noindent
\large
\textbf{Methods}\\
\normalsize

\noindent
\textbf{Full atom representation}  \\
\noindent
Each non-hydrogen atom present in the crambin crystal
structure \cite{CRAMBIN} (Brookhaven PDB accession code: 1AB1) was
represented by a hard sphere, whose size was given by scaling the
relevant VdW radius ($r$) from ref. \cite{CHOTHIA_VDW} by a factor
$\alpha (< 1)$.  Helix 1 and helix 2 native structures were obtained
by extracting residues 6-18 and 23-30, respectively, from the
crambin crystal structure.\\

\noindent
\textbf{Move set}\\
\noindent
A single MC step consisted of a backbone move followed by 10
sidechain moves.  Each backbone and sidechain move was accepted
according to the Metropolis criterion \cite{METROPOLIS}.  A backbone
move consisted of rotating the $\phi-\psi$ angles of up to 3
non-proline residues from a randomly selected window of 6 consecutive
residues.  A sidechain move consisted of rotating all sidechain
torsion angles ($\chi$) of a randomly selected non-proline residue.
The size of the backbone and sidechain rotations were obtained from a
Gaussian distribution with zero mean and standard deviation 2 and 10
degrees, respectively.\\

\noindent
\textbf{Square well G\={o} potential}\\
\noindent
We used an atomic square well potential \cite{MCQUARRIE} with the
well depths given by G\={o} energetics \cite{GO}.  In particular, for
two atoms $A$ and $B$ separated by a distance $R$, the energy
$\epsilon(A,B)$ was calculated according to

\[ \epsilon(A,B) = \left\{ 
\begin{array}{cc}
\infty & R < \sigma \\
\Delta(A,B) & \sigma \leq R <  \lambda\sigma \\
0 & R \geq \lambda\sigma \\
\end{array} \right.\ \ \ ,\]

\noindent
where $\sigma = \alpha (r_A + r_B)$ is the hard core distance,
$\lambda$ is a scaling factor $>1$, and $\Delta(A,B) = -1 $ if $A$ and
$B$ are in contact in the native conformation and 1 otherwise.  The
total energy of a conformation was computed as the sum over all pairs:
\[ E =\sum_{\textrm{all pairs}} \epsilon(A,B)\ \ \ .\]
All atom pairs of $i-i+1$ residues were excluded to eliminate any
biases towards local structure, and all backbone-backbone contacts
were ignored to eliminate non-specific interactions.  The energies of
the disulfide bonds were treated no differently from any other
contact.  We chose $\alpha = 0.75$ because it was the largest value
for which the native structure exhibited no steric clashes.
Furthermore, with $\alpha = 0.75$, we could not fold crambin with the
sidechain torsions held fixed at their native values, suggesting that
$\alpha$ was sufficiently large to enforce excluded volume
constraints.  The selection of small $\lambda$ values ($\leq 1.6$)
significantly increased the time of collapse, while large $\lambda$
values ($\geq 2.0$) made sidechain packing more degenerate.  We
therefore selected $\lambda = 1.8$ in order to balance the two
effects.  This makes the contact distance for methyl carbons to be
$5.08$ \AA.\\

\noindent
\textbf{Helix-coil transition thermodynamics and kinetics}\\
\noindent
Thermodynamic data were collected by sampling uncorrelated states
observed along long runs of at least $1.5 \times 10^{8}$ MC steps),
in order that multiple helix-colil transitions were observed at temperatures
near $T_{f}$.

Median first passage time data at a given temperature were collected
by performing 100 folding runs until the fraction of native contacts
$Q$ hit 0.7 or $10 \times 10^{6}$ MC steps had elapsed.  For the ensemble
kinetic data, 100 folding runs of length $5 \times 10^{6}$ were
collected, regardless of whether a folding event occurred.\\

\noindent
\textbf{Free energy calculations}\\
\noindent
Using thermodynamic runs, the average energy as a function of backbone
dRMS, $E(r)$, was first measured.  The dRMS of the uncorrelated states
was next histogrammed in bins of 0.2 \AA\ to compute the probability
of observing a particular dRMS, $p(r)$.  The entropy as a function of
backbone dRMS, $S = \ln W(r)$, was obtained by inverting the
statistical mechanical relation
\[ p(r) = \frac{W(r) e^{-\beta E(r)}}{Z} \]
where $W(r)$ is the density of states at a given dRMS and $Z$ is the
partition function \cite{FERRENBERG}.  The partition function was
explicitly determined from the $r=0$ bin by assuming that $W(r=0) \sim
1$.  Finally, the free energy at a temperature $T$ was obtained via
the identity $G(r) = E(r) - T S(r)$.\\

\noindent
\textbf{Crambin folding kinetics and thermodynamics}\\
\noindent
Random coils were first generated by unfolding crambin for $3 \times
10^{5}$ steps with only the excluded volume interaction turned on.
Both the average energy ($E = 77$, $Q = 0.02$) and average structure
($R_g = 19.5$, $\textrm{backbone dRMS} = 17.0$) of the random coils
indicate that these conformations are completely unfolded and
unstructured.  Each random coil was then simulated with the
square-well G\={o} potential turned on at a particular temperature
until it folded or $10^{8}$ MC steps elapsed.  Of the 250 runs
completed for $T = 1.25$ to $1.875$, 165 folded within our observation
window of $10^{8}$ steps.  For $1.4 \leq T \leq 1.8$, 135 out of 160
(= 84 \%) runs folded.  $2.5 \times 10^{6}$ MC
steps approximately took 1 hour of computation time on a Pentium III
550 Mhz PC.  \\

\noindent
\large
\textbf{Acknowledgements}\\
\normalsize
\noindent
We thank Henk Angerman, Gabriel Berriz, Michael Morrissey, Estelle
Pitard, and Juergen Wilder for helpful discussions.  This work was
supported by NIH grant R01-52126.

\newpage
\bibliographystyle{jmb} 
\bibliography{crambin_folding}

\newpage
\noindent
\large
\textbf{Figure captions}\\

\normalsize
\noindent
\underline{Figure 1} \\
\noindent
The 46-residue protein crambin.  The important secondary/tertiary
structural elements are indicated by different colors: black - helix 1
(residues 6-18, sequence: SIVARSNFNVCRL); red - helix 2 (23-30,
EAICATYT); green - inter-helix contacts; blue - $\beta$-sheet (1-5,
32-35, 38-46).  As shown by the matching colors, the $Q_{i}$
parameters are defined as the fraction of native contacts in specific
structural elements: $Q_{1}$ - helix 1; $Q_{2}$ - helix 2; $Q_{3}$ -
inter-helix contacts; $Q_{4}$ - $\beta$-sheet.\\
 
\noindent
\underline{Figure 2} \\
\noindent
Helix-coil transition trajectories (left panels) and energy histograms
(right panels) for helix 1 at various temperatures: above (T = 1.35;
top panels), near (T = 1.175; middle panels), and below (T = 0.9;
bottom panels) transition temperature.  \\

\noindent
\underline{Figure 3} \\
\noindent
Summary of helix-coil transition thermodynamics.  \textbf{a \& b}.
Transition curves for average E and backbone dRMS for helices 1 and 2,
respectively.  Note the difference in scales along the y-axis.
\textbf{c.}  Free energy of helix 1 formation at various temperatures
(see Methods).  The backbone dRMS is chosen as the order parameter.
Near the transition temperature (T = 1.175 - 1.225), two distinct free
energy minima appear (labeled ``helix'' and ``coil'').  \textbf{d.}
Heat capacity ($C_{v}$) curves for helices 1 and 2.  $C_{v}$ was
computed via the relation, $C_{v} = \left(\langle E^{2} \rangle-
\langle E \rangle^{2}\right)/T^{2}$, where $\langle\rangle$ refers to
ensemble averages \cite{MCQUARRIE}.\\

\noindent
\underline{Figure 4}\\
\noindent
Summary of helix-coil transition kinetics. \textbf{a}.  Median first
passage time (FPT) for helices 1 and 2.  \textbf{b-d}.  Ensemble
averaged kinetics of the early events during helix 1 formation at T =
0.7 and T = 1.175.  This represents the average fraction ($Q$) of
backbone amide (\textbf{b \& d}) and carbonyl (\textbf{c \& e}) native
contacts observed in 100 runs as a function of MC steps.  An increase
in the amide Q reflects helix formation towards the N-terminus, while
an increase in the carbonyl Q reflects helix formation towards the
C-terminus.  Note the difference in scales along the y-axes.\\

\noindent
\underline{Figure 5}\\
\noindent
The energy and heat capacity (shown as insert)
of crambin as a function of temperature ($T$).  The energy data
($\bullet$) were collected from uncorrelated structures sampled from
simulations of $10^{8}$ MC steps initiated from the native state.  For
$T < 2.0$, the energy data was fitted to a linear function ($R^{2} =
0.987$), while for $T \geq 2.0$, the exponential function $f(E,T) =
E_{n} + (E_{u} - E_{n})\exp(-C/T + D)/(1 + \exp(-C/T + D))$ ($C =
70.82$, $D = 34.16$, $E_{n} = $ native energy $= -658$, $E{u} = $
unfolded energy $= 16.67$; estimated $\sigma \approx 90$) was used.
The heat capacity ($C_{V}$) was obtained by evaluating $C_{V} = dE/dT$
on the fitted curves. We note that because our simulation does not
explicitly model solvent-protein interactions, the computed heat
capacity falls to zero for $T > T_f$, in contrast to experimental
observations.\\

\noindent
\underline{Figure 6}\\
\noindent
Typical folding runs at various temperatures.  The upper panel for
each run shows backbone (black) and sidechain (red) dRMS as the runs
progresses.  The lower panel tracks the four $Q_{i}$ values, with the
color coding shown in Figure \ref{fig:STABILITY}a.  Note that the
lengths are different for each run.  \textbf{a}.  Collapse-rate
limited cooperative folding at high temperature ($T = 1.875$).  The
secondary structure follows the sequence of events observed in the
fast-folding pathway (events 1-5 in Figure~\ref{fig:PATHWAY} are
labeled in the $Q_{i}$ plot).  \textbf{b}.  A trajectory ending at a
helix 2 misfold at low temperature ($T = 1.25$).  \textbf{c}.
Successful folding after encountering a helix 1 misfold at high
temperature ($T=1.775$).  This corresponds to pathway B in
Figure~\ref{fig:PATHWAY}. \textbf{d}.  Successful folding after
encountering a $\beta$-sheet misfold at high temperature ($T =
1.825$).  The $\beta$-sheet misfold is metastable at this temperature.
\textbf{e}.  A trajectory ending at a low temperature
sidechain-packing trap ($T = 1.425$).  \textbf{f}. Broken ergodicity
for $T < T^{*} \approx 1.6$.  For each temperature, 52 runs which
folded to near-native conformations (dRMS $< 1.25$ and $Q_{i}$'s $>
0.6$) were each extended for an additional $25 \times 10^{6}$ steps to
allow further relaxation.  The average of these extended runs are
indicated by $\circ$.  The average values obtained from simulations
started from the native state (see Figure \ref{fig:STABILITY}) are
indicated by $\bullet$.\\

\noindent
\underline{Figure 7}\\
\noindent
Summary of the folding kinetics.  The successive events
observed along the fast-folding pathway are marked by boxed numbers.
Note that the sidechain packing trap is not indicated on this
figure.\\

\noindent
\underline{Figure 8}\\
\noindent
Energy and backbone dRMS kinetics histograms
as a function of MC time of all runs at high ($1.7 < T < 1.8$; panel
\textbf{a}), middle ($1.5 < T < 1.7$; panel \textbf{b}), low ($ T <
1.5$; panel \textbf{c}) temperatures.  The color indicates the
fraction of runs at a particular energy or dRMS at a given MC time.
The color scale goes from blue (0.0) to red (0.1).  Because the runs
were terminated as they folded, after a run has folded it no longer
contributes to the histogram.  For this reason, as time progresses and
more runs fold, the color of the histograms moves towards blue.  At
all times, the histogram is normalized by the total number of runs at
step 0.  The data may be viewed as being obtained by a hypothetical
experiment that records the energy and dRMS of all non-folded structures
as the simulation progresses. The label ``hairpin'' refers to the
formation of inter-helix contacts.\\

\noindent
\underline{Figure 9}\\ Summary of the hairpin formation (residues
6-30) formation kinetics.  \textbf{a-c}.  Ensemble kinetics for
various temperatures.  The fraction of inter- and intra-helix native
contacts is plotted as a function of MC time.  The averages were taken
over 32 runs of $length 5.0 \times 10^{7}$.  \textbf{d \& e}.  A
typical folding trajectory at T = 0.4.  This trajectory exhibits
diffusion-collision behavior.\\

\newpage
%
%
%

\newpage
%
%
%
\newpage
%
%
%
\newpage
%
%
%

\newpage
%
%
%

\newpage
%
%
%

\newpage
%
%
%

\newpage
%
%
%

\newpage
%
%
%

\end{document}